# Optical Signatures of Strain Differences in Epitaxial Graphene Nanoribbons

Heather M. Hill[1], Ching-Chen Yeh[1,2], Swapnil M. Mhatre[1,3,4], Ngoc Thanh Mai Tran[1,5], Hanbyul Jin[1,5], Adam J. Biacchi[1], Chi-Te Liang[3], Angela R. Hight Walker[1], and Albert F. Rigosi[1*]

[1]*Physical Measurement Laboratory, National Institute of Standards and Technology (NIST), Gaithersburg, MD 20899, United States*

[2]*Graduate Institute of Applied Physics, National Taiwan University, Taipei 10617, Taiwan*

[3]*Department of Physics, National Taiwan University, Taipei 10617, Taiwan*

[4]*Department of Electrical, Computer & Energy Engineering, University of Colorado Boulder, Boulder, CO 80309, United States*

[5]*Joint Quantum Institute, University of Maryland, College Park, Maryland 20742, United States*

ABSTRACT: We demonstrate the preparation of both armchair and zigzag epitaxial graphene nanoribbons (GNRs) on 4H-SiC using a polymer-assisted, sublimation growth method. Historically, the preparation of GNRs depended on the quality, or smoothness, of the surface changes during growth. The physical phenomenon of terrace step formation introduces the risk of GNR deformation along sidewalls, but the risk is heavily mitigated by this polymer-assisted sublimation method. Two widths (100 nm and 50 nm) are examined electrically and optically for both armchair and zigzag GNRs. Our electrical results support the expected behaviors of the GNRs, while the optical signatures of variable strain reveal the subtle differences among all the GNR species measured.

[*] 100 Bureau Drive, Gaithersburg, MD 20899

# I. INTRODUCTION

Graphene nanoribbons (GNRs) are promising materials for future graphene-based nanoelectronics due to their unique band structure [1-3]. Theoretical calculations of the GNR band structure are well-explored and have established the substantial differences found between armchair (AC) GNRs [4-8], whose edge segments are rotated by ± 120° relative to the previous segment, and zigzag (ZZ) GNRs [9-12], whose edge segments alternate less abruptly with respect to the direction along the edge. Effects from quantum confinement are readily observable in GNRs [13-20], with their bandgap energies having an inversely proportional relationship to the GNR width. Despite this desirable quality for various field-effect transistor applications, engineering complications involving GNR width control, and by extension, band gap control, still remain. These problems warrant continued efforts to understand GNRs grown epitaxially.

In addition to the above difficulties, existing lithographic patterning methods typically cause rough and disordered edges to form during the etching process, possibly contributing to the degradation of the GNR's electrical properties [17-18, 21-29]. To overcome these drawbacks, structured silicon carbide (SiC) has been used as an ideal template for the selective growth of GNRs, which themselves have been shown to exhibit outstanding ballistic transport characteristics and electronic mean free paths of up to 15 μm [30-41]. Though using SiC alleviates many of the lithographic issues, problems that are exclusive to growths performed with SiC must be considered. In particular, the distortion of SiC surfaces and edges due to terrace formation has been shown to have substantial effects on GNR performance [28]. Although the effects of this phenomenon can be circumvented by fabricating devices from self-assembled GNRs on SiC [42], the total device sizes are restricted to the shorter of the two lateral terrace dimensions. For these reasons, improvements to the growth process are necessary. In recent work, polymer-assisted sublimation growth (PASG) techniques have enabled high quality graphene growth on centimeter scales due to the suppression of terrace formation [43-47]. These advances have improved graphene on SiC to the point that it has become a basis of fundamental research and development in resistance metrology [48-51].

In this work, the growths of approximately 100 nm-wide and 50 nm-wide GNR devices processed on 4H-SiC are demonstrated with improved structural and optical properties using PASG techniques. In this application, PASG promotes resistance on part of the slanted SiC sidewalls to deformation during the annealing process, allowing GNR growth to remain confined to the patterned SiC. The longitudinal magnetoresistances of both AC and ZZ GNR devices were measured at low temperatures, and the structural characteristics were examined by using atomic force microscopy (AFM), including the devices' conductive responses (C-AFM). Furthermore, Raman spectroscopic analyses were conducted to understand the variable nature of the influence of strain on the GNRs, with ZZ GNRs consistently showing greater strain when compared to the AC GNRs.

## II. EXPERIMENTAL METHODS

### A. GNR Growth and Device Fabrication

Square SiC chips were diced from on-axis 4H-SiC(0001) semi-insulating wafers (CREE) [see notes]. After cleaning the diced chips with piranha solution, SiC chips were submerged in a solution of hydrofluoric acid. Two etch processes were performed to shape the sidewalls on SiC prior to GNR growth (see Supplemental Material [52]). Chips were then processed with AZ5214E (a photoresist) for PASG [43]. The annealing process was performed with a graphite-lined resistive-element furnace (Materials Research Furnaces Inc.) [see notes]. The heating and cooling rates were about 1.5 °C/s, and the growth took place in an ambient argon environment at 1400 °C for 25 min, with GNRs forming along both the $(1\bar{1}00)$ and $(11\bar{2}0)$ crystallographic directions on SiC, representative of ZZ and AC orientations, respectively. Protective layers of Pd and Au were deposited on the GNRs to prevent organic contamination during the final etching of excess GNRs and deposition of contact pad metals (Pt and Ti). Excess GNRs were etched away to avoid interference with the intended devices. All further details are provided in the Supplemental Material [52].

### B. Atomic Force Microscopy

An Asylum Cypher [see notes] system was used in contact mode for AFM and C-AFM, with a Cr/Pt probe of radius 25 nm being used across the sample surface. The C-AFM scans were performed with a bias voltage of 10 mV applied to the sample. The setpoint varied between 0.2 V and 0.5 V and the gain parameter was set to 10. The GNRs' electrical conductivity and good adhesion to the substrate allowed for precise mapping of the nanostructures by C-AFM.

### C. Optical and Confocal Laser Scanning Microscopy

Optical microscopy was performed using a Nikon L200N optical microscope [see notes] in reflection mode using white light. Confocal laser scanning microscopy (CLSM) was performed using an Olympus LEXT OLS4100 system fitted with objectives ranging from 5× to 100× and provisions for an additional 8× optical zoom. The system utilizes a 405 nm wavelength semiconductor laser, which is scanned in the *x-y* directions by an electromagnetic micro-electro-mechanical systems (MEMS) scanner and a high-precision Galvano mirror. The use of this technique for graphene on SiC and similar materials has been well-described in other work [53]. The microscope was operated in an argon environment.

### D. Raman Spectroscopy

Raman spectroscopy was performed using a Renishaw InVia micro-Raman spectrometer [see notes] and a 633 nm wavelength excitation laser source. All spectra were measured and collected using a backscattering configuration with the sample upside-down to enhance the optical response from the GNR while lessening the influence of the optical response from the SiC substrate [54]. Other parameters include a 2 μm spot size, 240 s acquisition time, 1.7 mW power, 50 × objective, and 1800 mm$^{-1}$ grating. Each GNR sample was subject to a Raman line map measurement, which entailed collecting five or more points along each device, with lateral step sizes of about 1 μm.

## III. VERIFYING DEVICE FUNCTIONALITY

### A. Visual assessment of GNR growth with PASG

A basic illustration of the substrate etching process is shown in Fig. 1 (a). The primary flats of the full SiC wafers provide crystallographic orientation information to within 1°, which, in turn, provides guidance for the placement of sidewalls on the SiC chips. A CLSM image of the SiC substrate prior to growth can be seen in Fig. 1 (b). Once sidewalls were etched, the growth procedure was followed. The resulting device, along with an example AFM image of a GNR sidewall are shown in Fig. 1 (c) and (d), respectively.

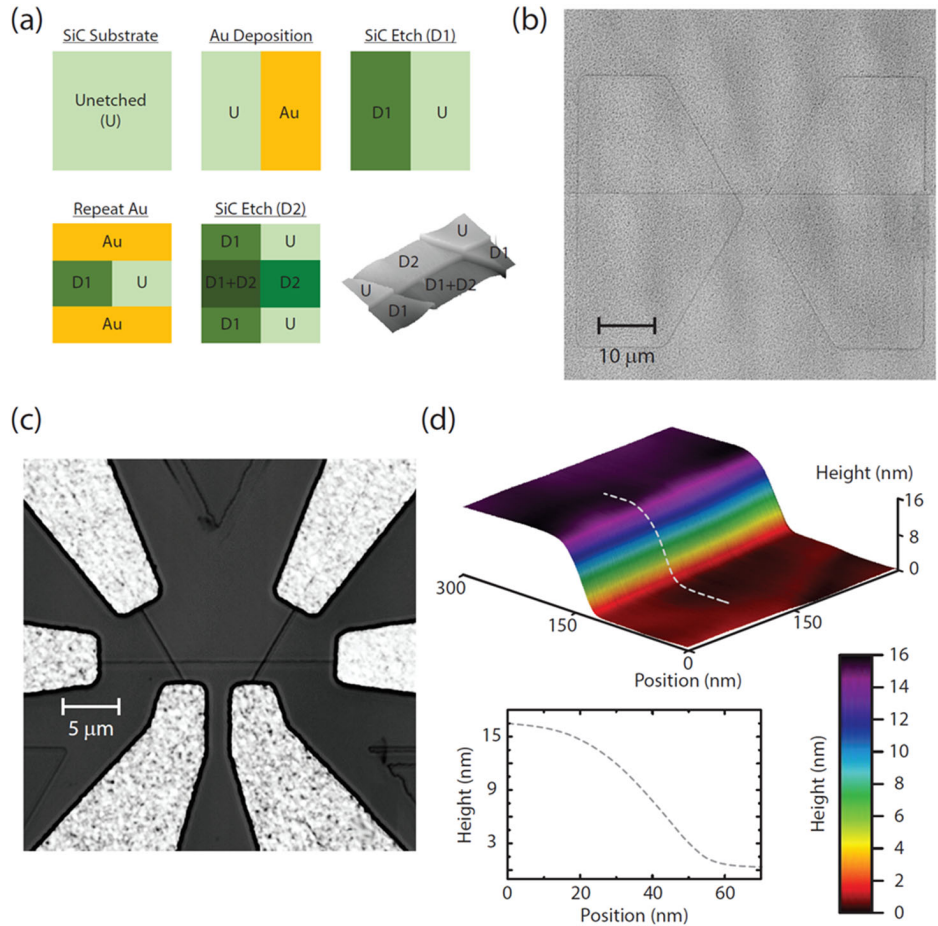

FIG. 1. (Color online) (a) An illustration of the SiC substrate etching process prior to GNR growth is shown. (b) A CLSM image is shown for an example substrate prior to growth. (c) A CLSM image depicts the final GNR Hall bar device. (d) A corresponding AFM image shows the quality of the GNR sidewall and adjacent surfaces.

### B. Conductive AFM of PASG-based GNRs

To assess whether the GNR devices behaved as usual AC or ZZ varieties, their electrical quality was inspected via C-AFM. Images from this technique were acquired on GNR devices of 50 nm width and both the AC and ZZ orientations were examined. In Fig. 2 (a) and (b), example C-AFM images show the ZZ and AC orientations, with the former exhibiting a higher conductivity by approximately two orders of magnitude (about a few nA versus some 10s of pA, respectively) [28]. At least a dozen sets of data were compiled into histograms in Fig. 2 (c) and (d), further supporting the trend that the ZZ orientation provided more conductive GNRs.

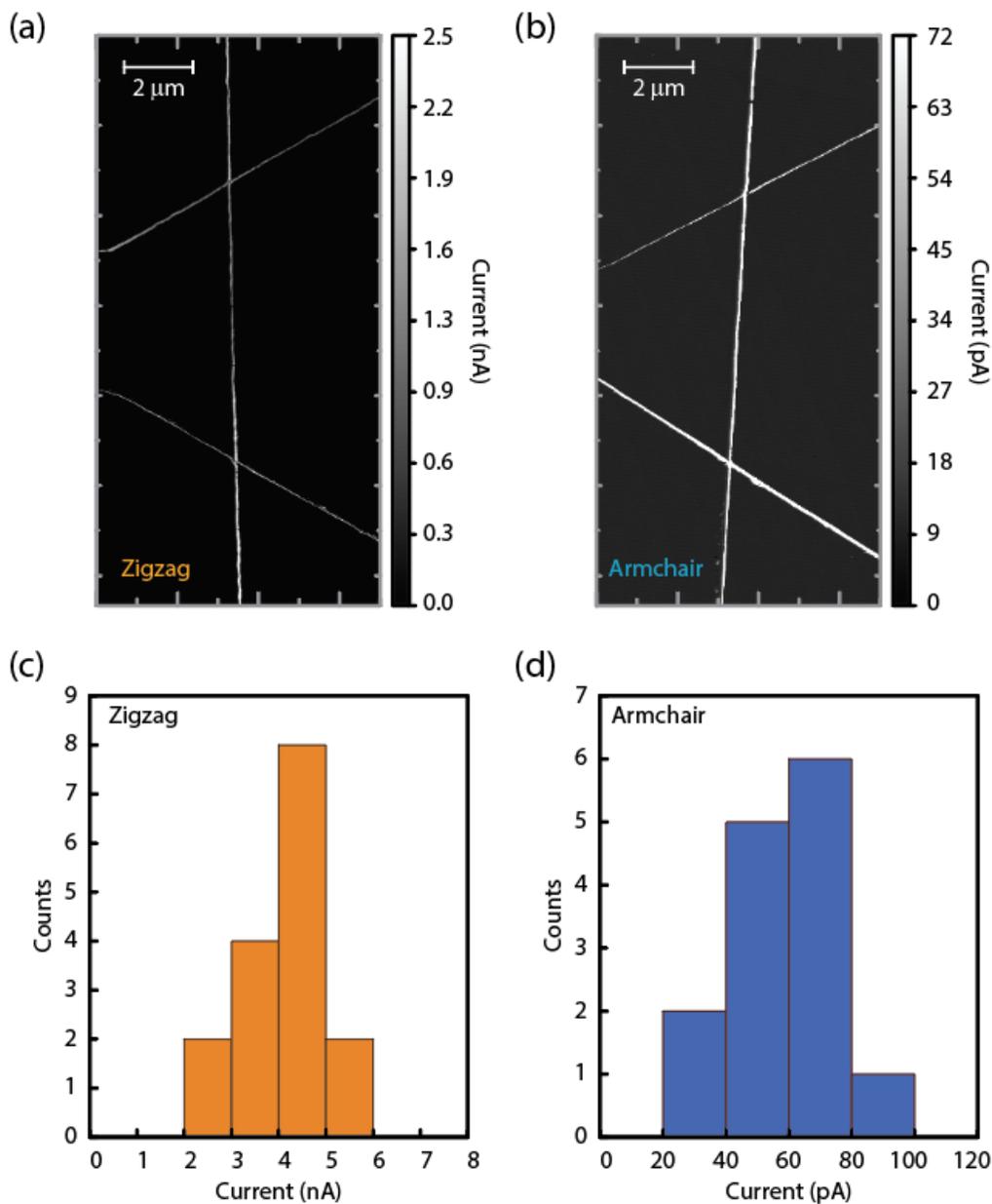

FIG. 2. (Color online) C-AFM images are shown for 50 nm-wide GNRs on 4H-SiC with the (a) ZZ orientation, which exhibits a higher conductivity than the (b) AC orientation. This technique is helpful in determining devices that would be unusable if some of its regions were not conductive at all, indicative of a lack of sufficient growth. (c) and (d) All counts are shown for a corresponding ZZ and AC GNR device, respectively. Note that the horizontal scales differ in order of magnitude.

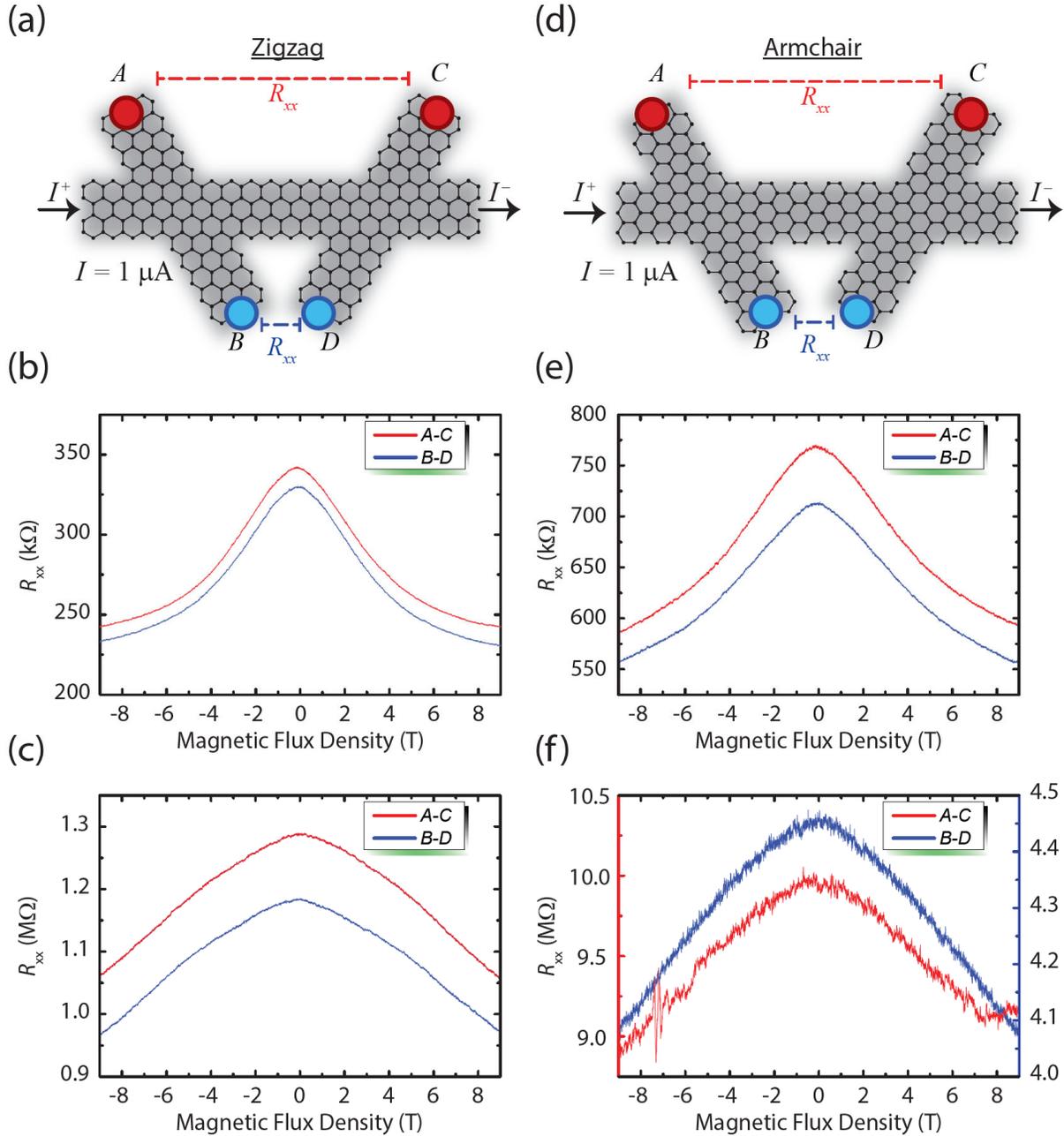

FIG. 3. (Color online) (a) Measurement configuration for the ZZ GNR. Electrical transport as a function of magnetic field was measured for the two widths: (b) 100 nm and (c) 50 nm. Similar measurement parameters were used for the (d) AC GNR configuration, yielding

magnetoelectrical transport data for the (e) 100 nm and (f) 50 nm-wide GNR devices. All data were collected at 1.5 K.

To avoid unnecessary introduction of errors into the later Raman experiments from devices that did not function properly, further verification was preferred. To show that devices were fully functional, magnetoelectrical transport measurements were performed using a current of 1 µA. Longitudinal resistance data were collected with an illustration of the measurement configurations shown in Fig. 3 (a) and (d) for ZZ and AC GNRs, respectively. For both widths, the measured ZZ resistances were always smaller than the AC GNRs by factors ranging from two to eight, which is an immediate reflection of the difference in band gap between the two species, with ZZ GNRs being more conductive generally [55].

## IV. RAMAN ANALYSES

Raman spectroscopy provides insight on the effects of strain on fully formed and functional GNR devices. Sets of example spectra are shown for the four GNR varieties in Fig. 4. From top to bottom, the panels display: 100 nm AC, 100 nm ZZ, 50 nm AC, and 50 nm ZZ. Though at least 30 spectra were acquired per variety, each of the Fig. 4 panels shows five example spectra that were collected from a line scan along the GNR, with a solid purple line representing the average which includes a 10-point adjacent-averaging smoothing. The 50 nm spectra have been multiplied by two to share a similar scale with the 100 nm data that use the same measurement parameters. There are five consistent modes that appear: D, D*, G-, G+, and 2D (G'). The first major observation is the emergence of a splitting in the G mode. This has historically been recognized as a manifestation of strain present in the GNRs [56-57]. Though doping also contributes to the shifting of various peak positions, the charge neutrality of epitaxially grown GNRs reduces this contribution substantially [35]. The second observation is the greater D peak presence in AC configurations regardless of ribbon width, or more specifically, a larger ratio of the intensity of the D mode when compared with the G mode. These two observations will be investigated more carefully in Fig. 5, and a more general, careful analysis of the peak fitting procedure will be presented in the coming paragraphs in the context of the Bayes factor.

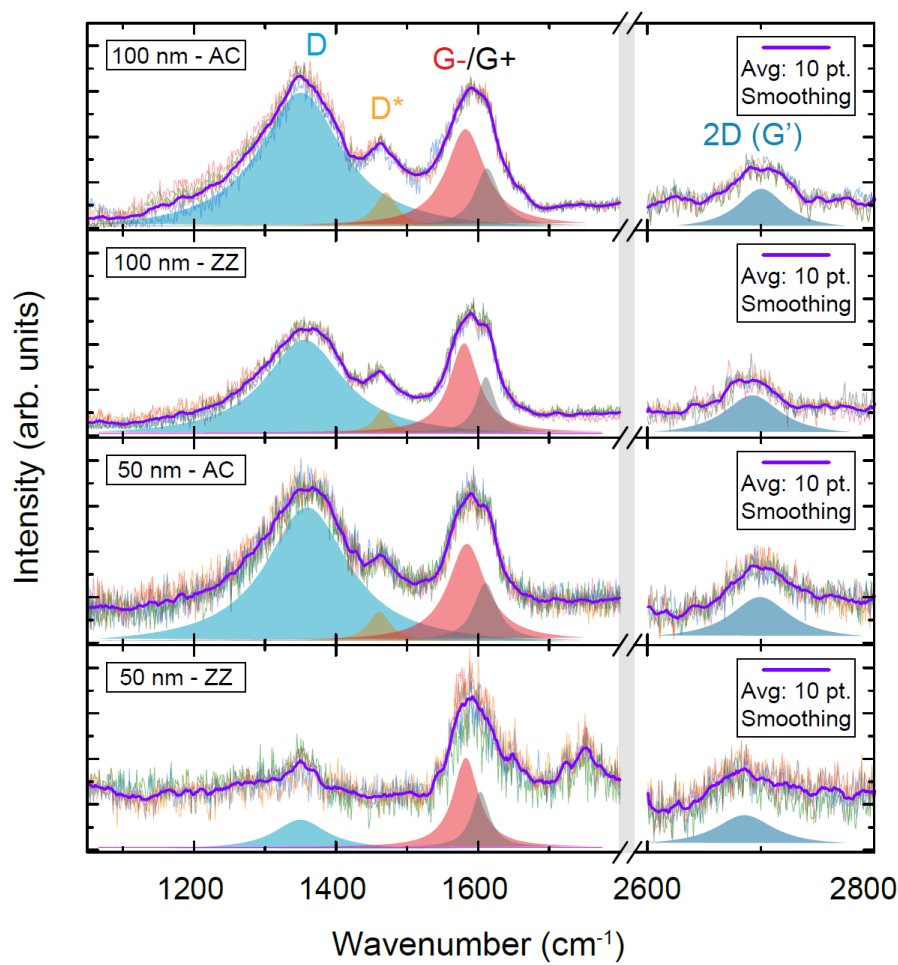

FIG. 4. (Color online) Sets of example spectra are shown for the four majority varieties of samples. From top to bottom, the panels display the following GNRs: 100 nm AC, 100 nm ZZ, 50 nm AC, and 50 nm ZZ. Each panel contains five examples taken along the GNR, with a solid purple line representing the average, with a 10-point adjacent-averaging smoothing applied. The 50 nm spectra have been multiplied by two to share a similar scale with the 100 nm data that use the same measurement parameters. The five generally consistent modes that appear are the D,

D*, G-, G+, and 2D (G'). Lorentzian profiles were used to fit these spectra to ultimately extract information about the D, G-, and G+ modes.

Before discussing the analysis related to strain contributions, it should be noted that in many spectra, a mode emerges in the mid-to-high 1400 cm$^{-1}$ region. There have been two possible explanations for this observation. The first, and more likely scenario, is the presence of edge vibrations for the AC and ZZ configurations, and these vibrations have been calculated to be a possible observation [58]. One limitation to this explanation would be the lack of observation of more prominent peaks for narrower GNR devices, as would be expected since there would be a weaker response from less material in the laser spot. The bottom two panels of Fig. 4 show spectra with weaker signals, and those signals appear noisier due to the applied scaling. Despite the similar measurement parameters, the observations of this reduction in response from the GNRs are primarily due to less material being excited (roughly half).

The second explanation that has been reported is that this peak can be attributed to a D* mode characteristic of the Si beneath as well as possible $sp^2$ contributions beneath the buffer layer (that are only bonded to the SiC substrate since amorphous carbon may result in broader peaks) [59-62]. This alternative explanation may be less likely due to a lack of observation of such bonds when buffer layer material was inspected by scanning transmission electron microscopy (combined with a high-angle annular dark field detector) [63].

In order to quantify the observations more carefully, Lorentzian profiles were used to fit all spectra to extract information about the D, G-, and G+ modes. And to justify the use of a two-peak model for all G-/G+ mode extractions, a numerical analysis was performed for each of three distinct fits to the experimental data – and more specifically, a single-, double-, and triple-peak model in the immediate neighborhood of the G mode. For more details on the fits, with some graphical examples, see the Supplemental Material [52]. It is through the calculation of the marginal likelihood integral (MLI) for each model that the Bayes factor, or ratio of MLIs, will better quantify the appropriateness of each of the fits [64, 65].

One must first define the MLI, where $n$ is the number of model parameters, $L_{max}$ is the maximum likelihood [65], $\mathbf{Cov_p}$ is the parameter covariance matrix, and $\Delta p$ is the parameter value range [64]:

$$MLI = (2\pi)^{n/2} L_{max} \frac{\sqrt{\det \mathbf{Cov_p}}}{\prod_{i=1}^{n} \Delta p_i}$$

(1)

As mentioned earlier, the MLI characterizes the appropriateness of a fit to its corresponding dataset, and the Bayes factor emerges when the ratio between the MLIs for two different models is taken [64]. Typically, when the Bayes factor is calculated, the quality of the model represented in the numerator of the ratio is being compared to the model in the denominator of the ratio. Therefore, the goal is to assess the two-peak model with respect to the one-peak model, bearing in mind that a Bayes factor over 100 (or, alternatively for large datasets, a logarithm of the Bayes factor greater than 5) indicates that the model of interest (two-peak model) is quantifiably more appropriate than the model to which it is compared (one-peak model). A Bayes factor closer to 1 or lower suggests that the model of interest is not very appropriate to use in lieu of its counterpart.

To provide an example evaluation of Eq. 1, the data in the second panel from the top in Fig. 4 (100 nm ZZ GNR data) are used in the neighborhood of the G mode (the green curve, approximately between 1520 cm$^{-1}$ and 1700 cm$^{-1}$, partly transparent) along with this two-peak fit:

$$y = y_0 + \frac{2A_1}{\pi} \frac{w_1}{4(x - x_{c1})^2 + w_1^2} + \frac{2A_2}{\pi} \frac{w_2}{4(x - x_{c2})^2 + w_2^2}$$

(2)

Several sets of relevant statistical quantities used in this analysis are provided the Supporting Material [52], including graphs of example fits and their covariance matrices. The corresponding covariance matrix output for this two-peak example is the following (where each column in this symmetric matrix is designated by the fitting parameters $y_0$, $x_{c1}$, $w_1$, $A_1$, $x_{c2}$, $w_2$, and $A_2$, respectively):

$$\begin{pmatrix} 6.6 \times 10^4 & -1.7 \times 10^4 & -3.5 \times 10^3 & -2.9 \times 10^8 & -5.1 \times 10^3 & 1.6 \times 10^4 & 2.6 \times 10^8 \\ -1.7 \times 10^4 & 5.6 \times 10^3 & 5.4 \times 10^2 & 9.6 \times 10^7 & 1.8 \times 10^3 & -5.6 \times 10^3 & -8.8 \times 10^7 \\ -3.5 \times 10^3 & 5.4 \times 10^2 & 3.9 \times 10^2 & 8.8 \times 10^6 & 81 & -4.4 \times 10^2 & -7.4 \times 10^6 \\ -2.9 \times 10^8 & 9.6 \times 10^7 & 8.8 \times 10^6 & 1.6 \times 10^{12} & 3.2 \times 10^7 & -9.7 \times 10^7 & -1.5 \times 10^{12} \\ -5.1 \times 10^3 & 1.8 \times 10^3 & 81 & 3.2 \times 10^7 & 6.5 \times 10^2 & -1.9 \times 10^3 & -3.0 \times 10^7 \\ 1.6 \times 10^4 & -5.6 \times 10^3 & -4.4 \times 10^2 & -9.7 \times 10^7 & -1.9 \times 10^3 & 5.7 \times 10^3 & 8.9 \times 10^7 \\ 2.6 \times 10^8 & -8.8 \times 10^7 & -7.4 \times 10^6 & -1.5 \times 10^{12} & -3.0 \times 10^7 & 8.9 \times 10^7 & 1.4 \times 10^{12} \end{pmatrix}$$

Above, $\sqrt{det\ \mathbf{Cov_p}}$ can be calculated for this two-peak fit, and the value is on the order of $10^{12}$, $n$ is 7, and $\prod_{i=1}^{n} \Delta p_i$ is approximately on the order of $10^{20}$. The latter is based on the fit parameters ($y_0$, $x_{cn}$, $w_n$, $A_n$) having ranges of $10^2$, $10^2$, $10^1$, and $10^6$, respectively. The units for $\sqrt{det\ \mathbf{Cov_p}}$ will vary based on the number of peaks in the model. For calculating $L_{max}$, the means outlined in Ref. [64] were followed, with the most important feature being the determination of the probability distribution function of the dataset. Due to the large value of this computation, it is more helpful to compute the logarithm of the value, resulting in $log\ (L_{max})$ being about 1672 for the two-peak, ZZ GNR model.

This rigorous analysis was repeated for the one- and three-peak model for each dataset, but the latter, as well as the one-, two-, and three-peak analysis for an exemplary AC GNR dataset in the G mode neighborhood, can be found in the Supplemental Material [52]. For the ZZ GNR G mode one-peak model, $\sqrt{det\ \mathbf{Cov_p}}$ was calculated to be about $10^5$, $n$ is 4, $L_{max}$ is about 1456, and $\prod_{i=1}^{n} \Delta p_i$ is on the order of $10^{11}$. From these factors, the *logarithm* of the Bayes factor comparing the two-peak model to the one-peak model is much greater than 5 (with the Bayes factor itself much greater than 100), thus indicating that the two-peak model is a decisively stronger model to use.

Regarding the questions of whether a three-peak model would be more appropriate for the G mode neighborhood, the full Bayes factor analysis was repeated with the two-peak model in the denominator of the Bayes factor ratio. In the three-peak model, $\sqrt{det\ \mathbf{Cov_p}}$ was calculated to be about $6 \times 10^{10}$, $n$ is 10, $L_{max}$ is about 1662, and $\prod_{i=1}^{n} \Delta p_i$ is on the order of $10^{29}$. From these factors, the *logarithm* of the Bayes factor comparing the three-peak model to the two-peak model was a positive number much smaller than 1 (and a logarithm on the order of –10), meaning that the three additional parameters introduced by the third peak were not appropriate from a statistical perspective. More information on this analysis, including analyses on the D and 2D modes, can be found in the Supplemental Material [52].

The positions and widths of the final Lorentzian fits are presented in Fig. 5 in the following order: (a) 100 nm AC, (b) 100 nm ZZ, (c) 50 nm AC, and (d) 50 nm ZZ. In all panels, the dashed lines indicate the average value, surrounded by a shaded region of 1σ uncertainty associated with the averaging process. In all the panels for the split G mode, the long-dashed vertical and horizontal averages (red and orange, respectively) are associated with the G- mode, whereas the short-dashed vertical and horizontal averages (black and blue, respectively) are associated with the G+ modes. All bottom panels correspond to the D mode, where the long-dashed lines surrounded by orange (position) and red (FWHM) shading again give the average values with 1σ uncertainties associated with the averaging process.

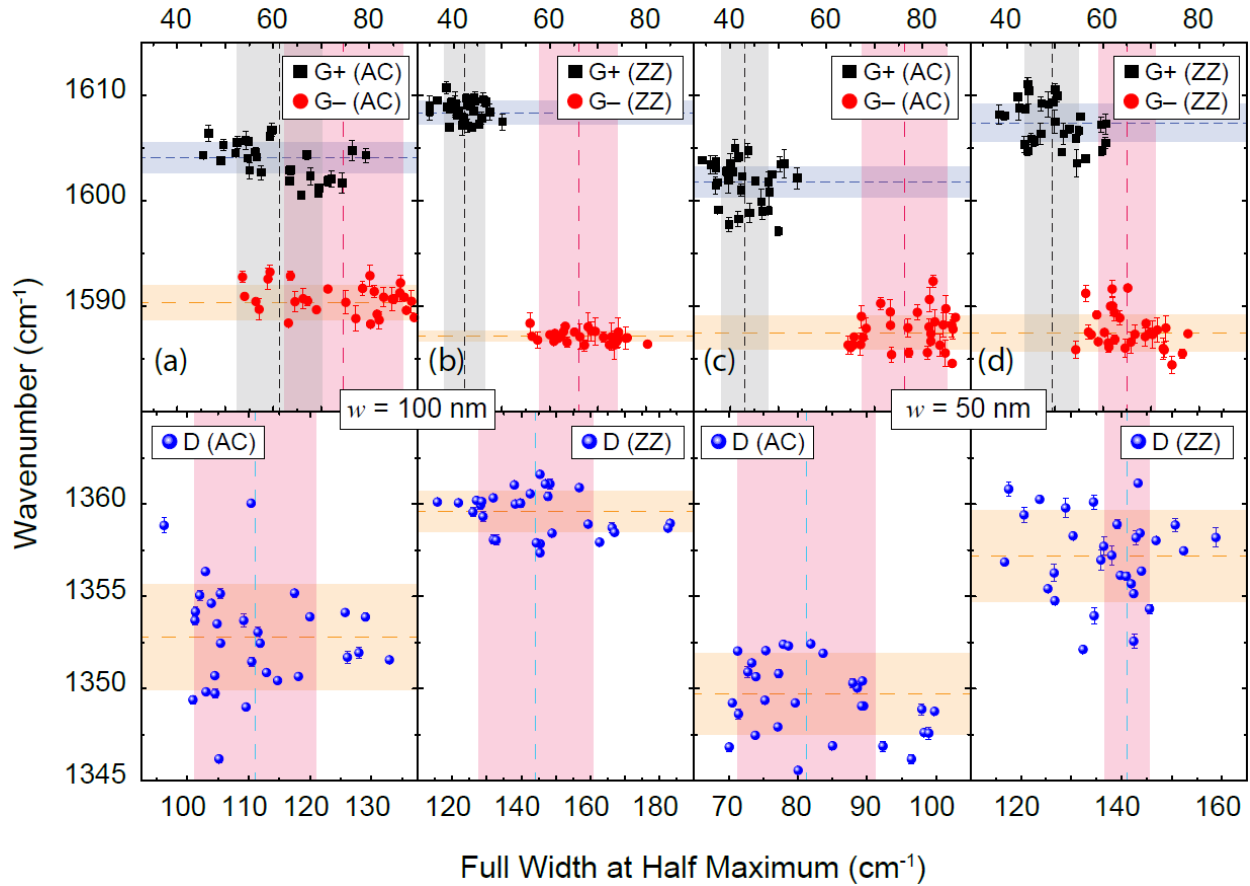

FIG. 5. (Color online) The GNR spectra are analyzed to extract each position and FWHM of the fitted D, G-, and G+ modes, with each pair of vertically aligned panels designating: (a) 100 nm AC, (b) 100 nm ZZ, (c) 50 nm AC, and (d) 50 nm ZZ. In all panels, the dashed lines indicate the average value, surrounded by a similarly colored shading indicating the region of 1σ uncertainty associated with the averaging process. In all top panels, the long-dashed vertical and horizontal (red and orange, respectively) averages are associated with the G- modes, whereas the short-dashed averages (black and blue for vertical and horizontal, respectively) are associated with the G+ modes. All bottom panels correspond to the D mode. All data points have error bars (some of which are smaller than the points) indicating the 1σ uncertainty associated with the fitting procedure.

The first inference one can make from the extracted splitting of the G mode in Fig. 5 is that the ZZ GNRs experience greater strain than their AC counterparts, as can be seen by the splitting separations of 21.4 cm$^{-1}$ ± 1.5 cm$^{-1}$ and 19.6 cm$^{-1}$ ± 3.8 cm$^{-1}$ (for 100 nm and 50 nm ZZ GNRs, respectively) versus the separations of 13.3 cm$^{-1}$ ± 3.3 cm$^{-1}$ and 13.9 cm$^{-1}$ ± 3.9 cm$^{-1}$ (for 100 nm and 50 nm AC GNRs, respectively). Mohiuddin *et al*. found that the separation of this doubly degenerate mode scaled with strain [56]. Based on this work, the corresponding strains for the ZZ and AC GNRs would be approximately 1.1 % and 0.8 %, respectively. The key difference of 0.3 % is the notable takeaway from these measurements.

An additional support for this first inference and key takeaway in strain difference comes from the relative shift of the GNR's 2D (G') mode. As mentioned earlier, information on the Bayes factor analysis for the 2D mode can be found in the Supplemental Material [52], and this information allows one to justify tracking one-peak behavior rather than two-peak behavior. The average values of the 2D mode position are: 2699.8 cm$^{-1}$ ± 2.4 cm$^{-1}$ (AC GNR, 50 nm), 2700.1 cm$^{-1}$ ± 1.8 cm$^{-1}$ (AC GNR, 100 nm), 2685.1 cm$^{-1}$ ± 1.6 cm$^{-1}$ (ZZ GNR, 50 nm), and 2689.5 cm$^{-1}$ ± 2.6 cm$^{-1}$ (ZZ GNR, 100 nm). From what has been reported in a GNR experiencing strain [56-57], in addition to the redshifting of the 2D mode in ZZ-GNRs, this approximately 10 cm$^{-1}$ to 15 cm$^{-1}$ difference in wavenumber between AC and ZZ GNRs corresponds to a similar difference in exhibited strain (about 0.3 %). Since the differences in strain determined by the G and 2D modes are nearly identical, it is not unreasonable to neglect additional contributions to mode shifting.

The second inference that can be made regards the direction of this new strain. The G+ and G- modes represent orthogonal phonon eigenvectors, and other works have reported that these modes can soften, or redshift, more strongly based on the directionality of the newly applied, uniaxial strain [56, 66]. More specifically, the G+ mode is perpendicular to the applied strain and the G- mode is parallel to the applied strain [67-70]. It has been observed that when growing GNRs on SiC, additional strain is introduced along the direction of the ribbon itself [71]. If one revisits Fig. 5, one can see that there is greater shifting observed in the G+ mode of the ZZ GNRs. Because the scanning tunneling microscope images in Ref. [71] show strain in AC GNRs as along the ribbon, it then follows that the extra strain experienced by the ZZ GNRs more likely to occur perpendicular to the GNR device rather than along its length.

It will be important to better correlate specific changes in strain to corresponding changes in the band structures of these sidewall materials, and subsequently, learn more about those modified band structures' ramifications on expected device electrical behaviors. Overall, these observations and analyses are necessary for building GNR devices, especially if those devices are aimed at harboring ballistic transport.

## V. CONCLUSIONS

In this work, growths of GNR devices about 100 nm and 50 nm wide on 4H-SiC have been demonstrated using PASG techniques. In this application, PASG promotes the deformation resistance of the slanted SiC sidewalls during the annealing process, allowing GNR growth to remain confined to the patterned SiC. The longitudinal magnetoresistances of both AC and ZZ GNR devices were measured at low temperatures in conjunction with C-AFM to verify device functionality. Upon confirmation, devices were tested optically with Raman spectroscopy and data analyses were conducted to understand the variable nature of the influence of strain on the GNRs, with ZZ GNRs consistently showing a greater experience of strain when compared to the AC GNRs. The knowledge obtained from the optical and electrical characterization will be applicable to future GNR device fabrication and especially relevant to those applications seeking to utilize ballistic transport.

## ACKNOWLEDGMENTS AND NOTES

The work of CCY and SMM at NIST was made possible by arrangement with C.-T. Liang of National Taiwan University. The authors would like to thank L. Chao, M. L. Kelley, G. Fitzpatrick, and E. C. Benck for their assistance in the NIST internal review process. Work presented herein was performed, for a subset of the authors, as part of their official duties for the United States Government. Funding is hence appropriated by the United States Congress directly. The authors declare no competing interest.

Commercial equipment, instruments, and materials are identified in this paper in order to specify the experimental procedure adequately. Such identification is not intended to imply recommendation or endorsement by the National Institute of Standards and Technology or the United States Government, nor is it intended to imply that the materials or equipment identified are necessarily the best available for the purpose.

Supplemental Material

# Optical Signatures of Strain Differences in Epitaxial Graphene Nanoribbons

Heather M. Hill[1], Ching-Chen Yeh[1,2], Swapnil M. Mhatre[1,3,4], Ngoc Thanh Mai Tran[1,5], Hanbyul Jin[1,5], Adam J. Biacchi[1], Chi-Te Liang[3], Angela R. Hight Walker[1], and Albert F. Rigosi[1†]

[1]*Physical Measurement Laboratory, National Institute of Standards and Technology (NIST), Gaithersburg, MD 20899, United States*

[2]*Graduate Institute of Applied Physics, National Taiwan University, Taipei 10617, Taiwan*

[3]*Department of Physics, National Taiwan University, Taipei 10617, Taiwan*

[4]*Department of Electrical, Computer & Energy Engineering, University of Colorado Boulder, Boulder, CO 80309, United States*

[5]*Joint Quantum Institute, University of Maryland, College Park, Maryland 20742, United States*

**GNR growth and fabrication procedures**

The details of the SiC substrate preparation, GNR growth, and additional fabrication steps are provided here.

The steps below are for <u>SiC etching</u> to create sidewalls for eventual GNR growth:

1. **Spin coating on bare SiC**

    a. LOR3A photoresist spun at 418.88 rad/s for 45 s

    b. Substrate baked at 180 °C for 5 min

    c. S1813 photoresist spun at 471.23 rad/s for 45 s

    d. Substrate baked at 115 °C for 1 min

---

[†] Email: albert.rigosi@nist.gov

2. **UV Exposure**

    a. Use photolithography tool MLA150 at 375 nm with 90 mJ/cm$^2$

3. **Developing**

    a. CD 26A for 50 sec, followed by rinse with deionized water

4. **Metal deposition with e-beam deposition**

    a. Pd and Au (total respective thicknesses 10 nm and 150 nm) (respective deposition rates of 0.05 nm/s and 0.15 nm/s)

    b. Lift-off with PG remover at 80 °C for 1 h

    c. Repeat with fresh PG remover for 5 min

5. **Dry etching of SiC**

    a. Reactive ion etching tool used with SF$_6$:O$_2$ (both gases have flow rate of 20 cm³/min at standard temperature and pressure), 13.33 Pa, and 100 W for 90 s

    b. Note: this etches approximately 20 nm of SiC

6. **Removing etching mask**

    a. Place small patterns using S1813 to establish alignment marks for later steps

    b. Etch Au with a diluted solution of aqua regia and deionized water (1:1) for 2 min

    c. Remove photoresist S1813 with PG remover (80 °C for 10 min)

    d. Rinse substrate with isopropyl alcohol

The PASG technique and preparation for <u>growth</u> is summarized here:

1. **Sample cleaning**

    a. Clean substrate with acetone, then isopropyl alcohol – each step is 10 min and includes sonication

    b. Place substrate in piranha bath at 120 °C for 10 min

    c. Hydrofluoric (HF) acid bath (49 % concentration by volume) for 2 min (Step 2b immediately after)

2. **Spin-coating of polymer for growth assistance**

    a. Solution: 150 mL of isopropyl alcohol and 30 droplets (approximately 200 μL) AZ5214E

    b. Spin solution at 628.32 rad/s (with initial acceleration of 418.88 rad/s$^2$) for 30 s total process immediately after HF cleaning

3. **GNR growth (facing polished graphite disk)**

    a. 1200 °C for 30 min to induce sidewall sloping

    b. 1400 °C for 30 min to induce growth of GNRs

To <u>remove excess GNRs</u> outside the desired Hall bar region:

1. **Metal deposition for protective layer**

    a. Pd and Au (total respective thicknesses 10 nm and 30 nm) (respective deposition rates of 0.05 nm/s and 0.1 nm/s)

2. **Spin coating on bare SiC**

    a. LOR3A photoresist spun at 418.88 rad/s for 45 s

    b. Substrate baked at 180 °C for 5 min

    c. S1813 photoresist spun at 471.23 rad/s for 45 s

    d. Substrate baked at 115 °C for 1 min

3. **UV Exposure**

    a. Use photolithography tool MLA150 at 375 nm with 90 mJ/cm$^2$

4. **Developing**

    a. CD 26A for 50 sec, followed by rinse with deionized water

5. **Metal deposition with e-beam deposition**

    a. Cr (70 nm total thickness) (0.05 nm/s and 0.03 nm/s)

    b. Lift-off with PG remover at 80 °C for 1 h

    c. Repeat with fresh PG remover for 5 min

6. **Dry etching of remaining Pd/Au and excess GNRs**

    a. Reactive ion etching tool used with Ar (40 cm$^3$/min at standard temperature and pressure), 4 Pa, and 150 W for 10 min, then repeat for 3 min

7. **Removing etching mask**

    a. Use Cr etchant 1030 for 3 min

The next step places <u>electrical contact pads</u> on the GNR devices:

1. **Spin coating on bare SiC**

    a. LOR3A photoresist spun at 418.88 rad/s for 45 s

    b. Substrate baked at 180 °C for 5 min

    c. S1813 photoresist spun at 471.23 rad/s for 45 s

    d. Substrate baked at 115 °C for 1 min

2. **UV Exposure**

    a. Use photolithography tool MLA150 at 375 nm with 90 mJ/cm$^2$

3. **Developing**

    a. CD 26A for 50 sec, followed by rinse with deionized water

4. **Metal deposition with e-beam deposition**

    a. Pd and Au (total respective thicknesses 10 nm and 200 nm) (respective deposition rates of 0.05 nm/s and 0.15 nm/s)

    b. Lift-off with PG remover at 80 °C for 1 h

    c. Repeat with fresh PG remover for 5 min

   d. Note: this leaves part of the electrical contacts formed. The contact pads for wire bonding are listed next.

5. **Spin coating on bare SiC**

   a. LOR3A photoresist spun at 418.88 rad/s for 45 s

   b. Substrate baked at 180 °C for 5 min

   c. S1813 photoresist spun at 471.23 rad/s for 45 s

   d. Substrate baked at 115 °C for 1 min

6. **UV Exposure**

   a. Use photolithography tool MLA150 at 375 nm with 90 mJ/cm$^2$

7. **Developing**

   a. CD 26A for 50 sec, followed by rinse with deionized water

8. **Wire bonding metal deposition**

   a. Ti and Pt (total respective thicknesses 10 nm and 250 nm) (respective deposition rates of 0.05 nm/s and 0.1 nm/s)

   b. Lift-off with PG remover at 80 °C for 1 h

Repeat with fresh PG remover for 5 min.

**Bayes Factor Analysis for Raman Spectra**

Additional Bayes factor analyses were performed to ensure the appropriateness of using a two-peak model for the G mode neighborhood. As such, all graphs presented herein have Counts as a vertical axis and Wavenumber (in cm$^{-1}$) as the horizontal axis.

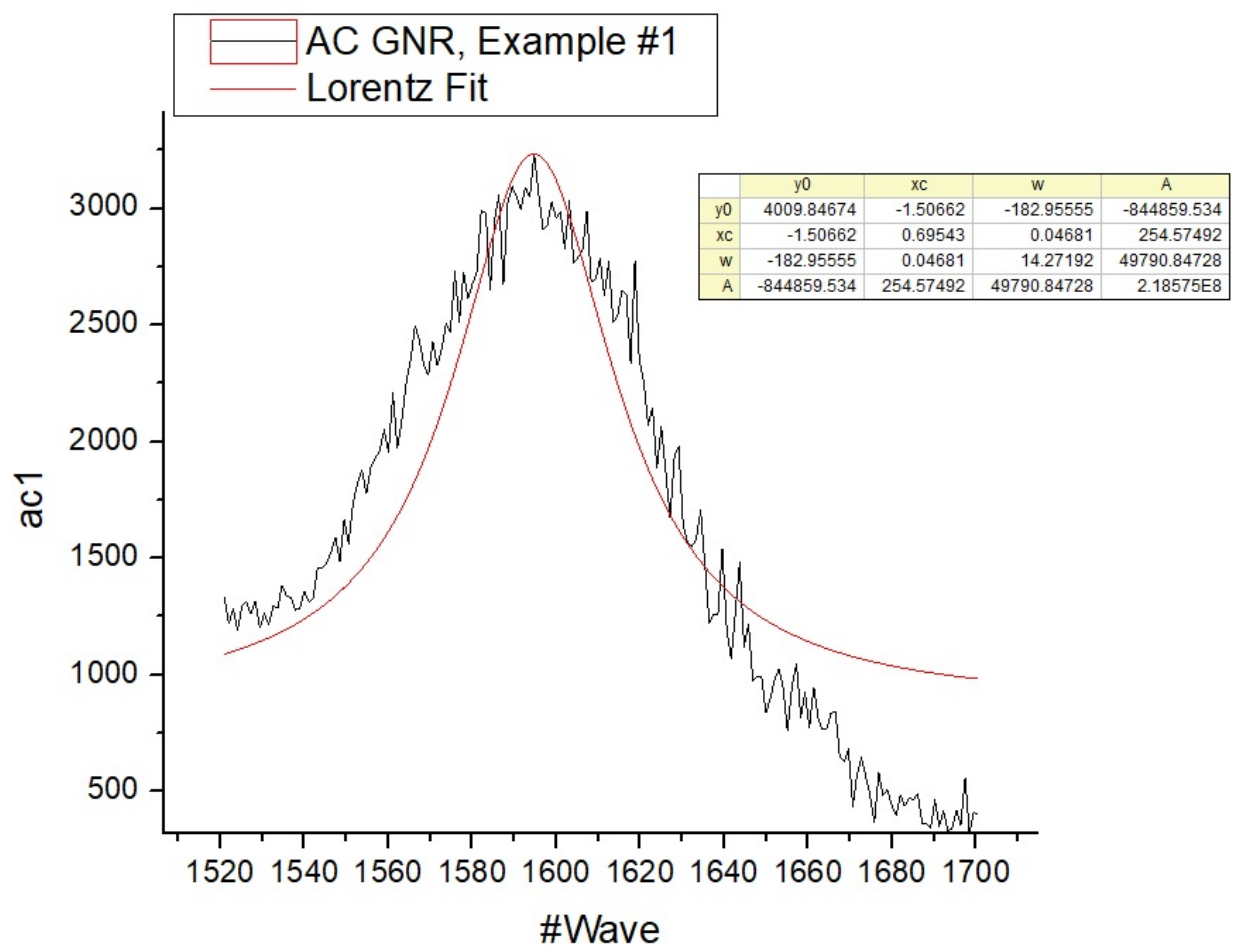

**Fig. 1-SM**. AC GNR example Raman spectrum (100 nm device) with one-peak Lorentzian fit. Each fit had its covariance matrix calculated, and in this case, may be seen at the top right corner of the graph.

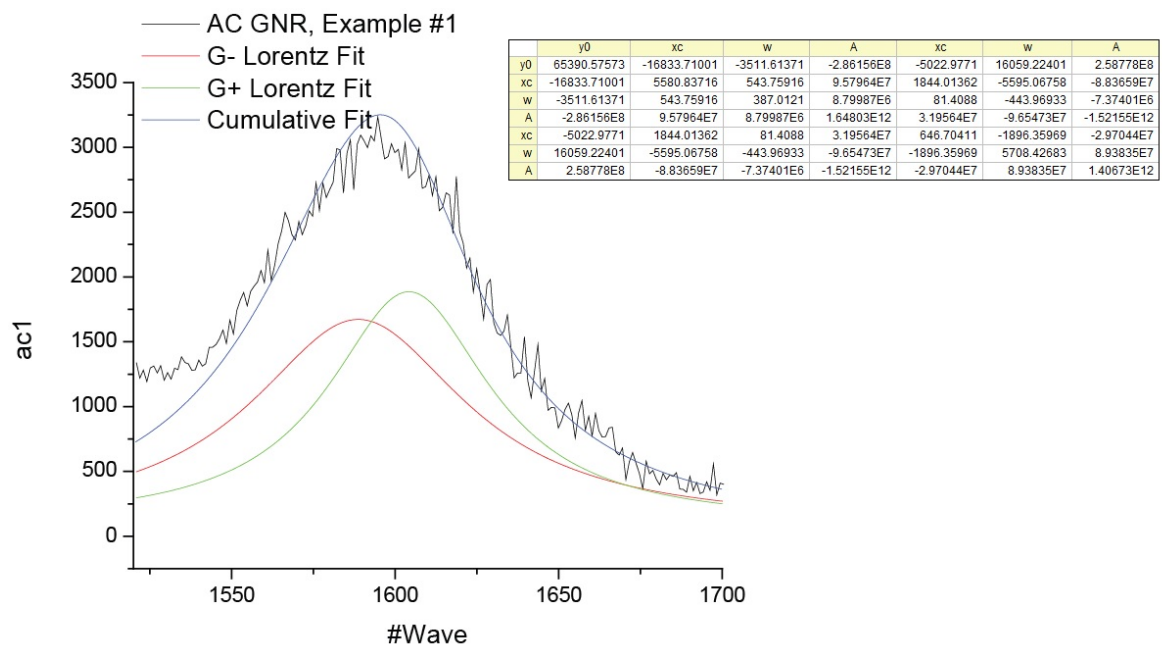

**Fig. 2-SM**. AC GNR example Raman spectrum (100 nm device) with two-peak Lorentzian fit. The covariance matrix may be seen at the top right corner of the graph.

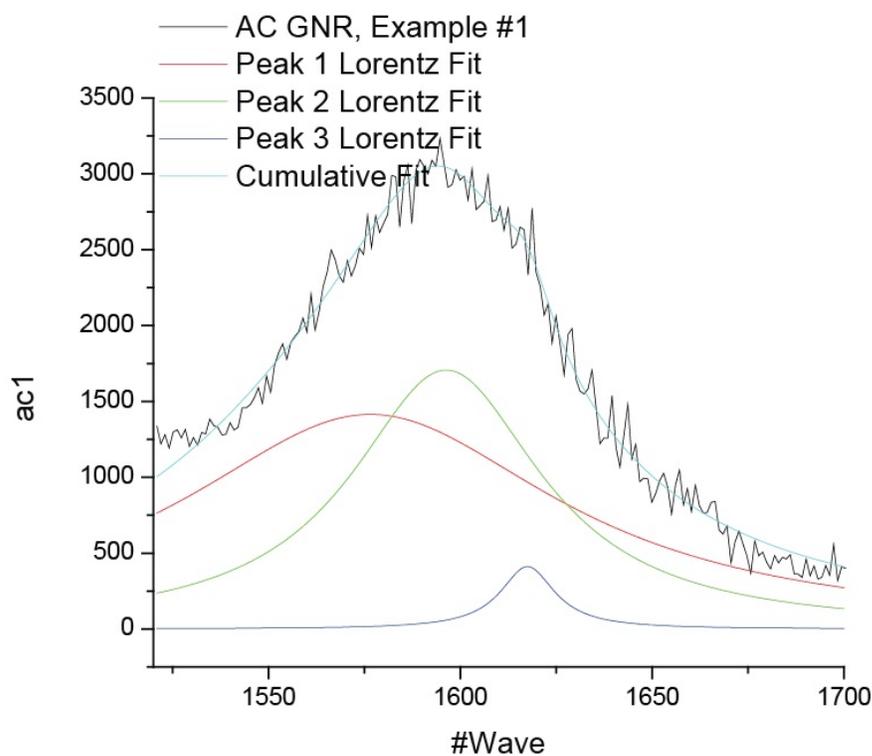

|    | y0          | xc           | w           | A            | xc            | w            | A            | xc            | w            | A            |
|----|-------------|--------------|-------------|--------------|---------------|--------------|--------------|---------------|--------------|--------------|
| y0 | 25106.31308 | -2869.09963  | -994.7859   | -3.89351E7   | 236.06499     | 3702.58218   | 2.86864E7    | 98.68167      | -989.53718   | -1.45645E6   |
| xc | -2869.09963 | 435.70695    | -5.95075    | 5.63808E6    | -26.13329     | -586.56278   | -4.5227E6    | -11.33912     | 166.26087    | 236744.55504 |
| w  | -994.7859   | -5.95075     | 246.19364   | 278885.28888 | -28.06498     | 21.91112     | 212023.15556 | -7.57905      | 12.63595     | 20471.63425  |
| A  | -3.89351E7  | 5.63808E6    | 278885.28888| 7.38988E10   | -363543.54731 | -7.62139E6   | -5.85818E10  | -150636.64824 | 2.16696E6    | 3.08862E9    |
| xc | 236.06499   | -26.13329    | -28.06498   | -363543.54731| 8.01057       | 35.89908     | 268930.44962 | 2.35103       | -21.98656    | -28353.56385 |
| w  | 3702.58218  | -586.56278   | 21.91112    | -7.62139E6   | 35.89908      | 810.82032    | 6.19251E6    | 13.73284      | -238.28756   | -335199.18856|
| A  | 2.86864E7   | -4.5227E6    | 212023.15556| -5.85818E10  | 268930.44962  | 6.19251E6    | 4.75377E10   | 111760.52875  | -1.81647E6   | -2.55285E9   |
| xc | 98.68167    | -11.33912    | -7.57905    | -150636.64824| 2.35103       | 13.73284     | 111760.52875 | 2.05924       | -7.61738     | -9669.08387  |
| w  | -989.53718  | 166.26087    | 12.63595    | 2.16696E6    | -21.98656     | -238.28756   | -1.81647E6   | -7.61738      | 117.60678    | 142073.04882 |
| A  | -1.45645E6  | 236744.55504 | 20471.63425 | 3.08862E9    | -28353.56385  | -335199.18856| -2.55285E9   | -9669.08387   | 142073.04882 | 1.83747E8    |

**Fig. 3-SM**. AC GNR example Raman spectrum (100 nm device) with three-peak Lorentzian fit. The covariance matrix may be seen beneath the graph.

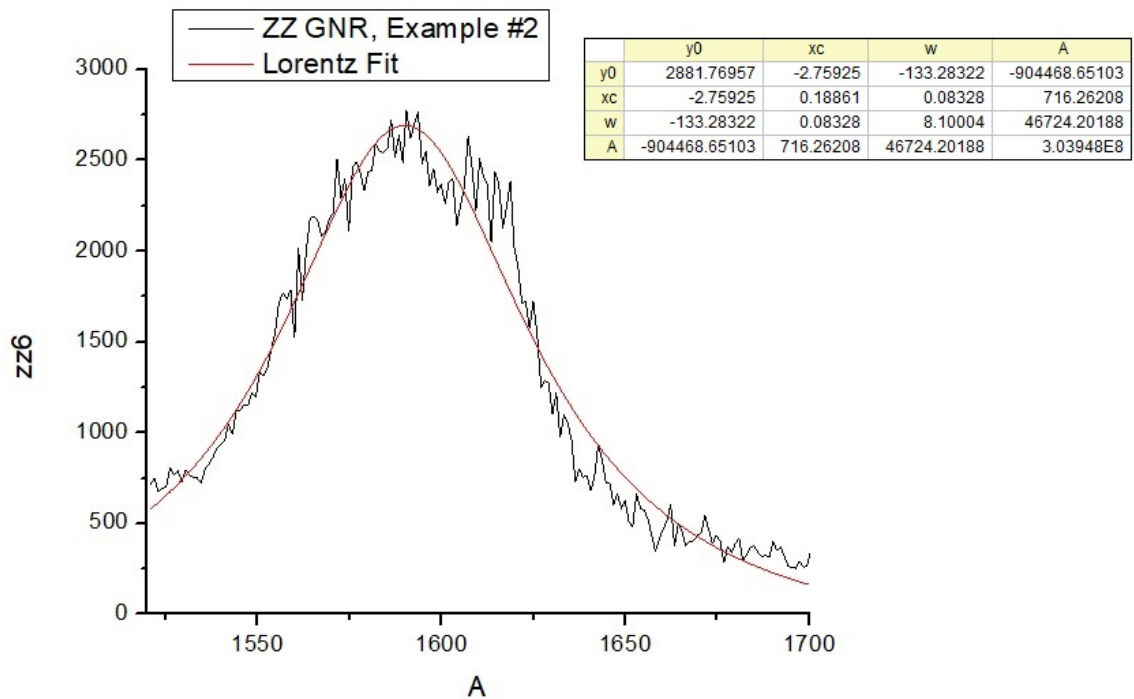

**Fig. 4-SM**. ZZ GNR example Raman spectrum (100 nm device) with one-peak Lorentzian fit. The covariance matrix may be seen at the top right corner of the graph.

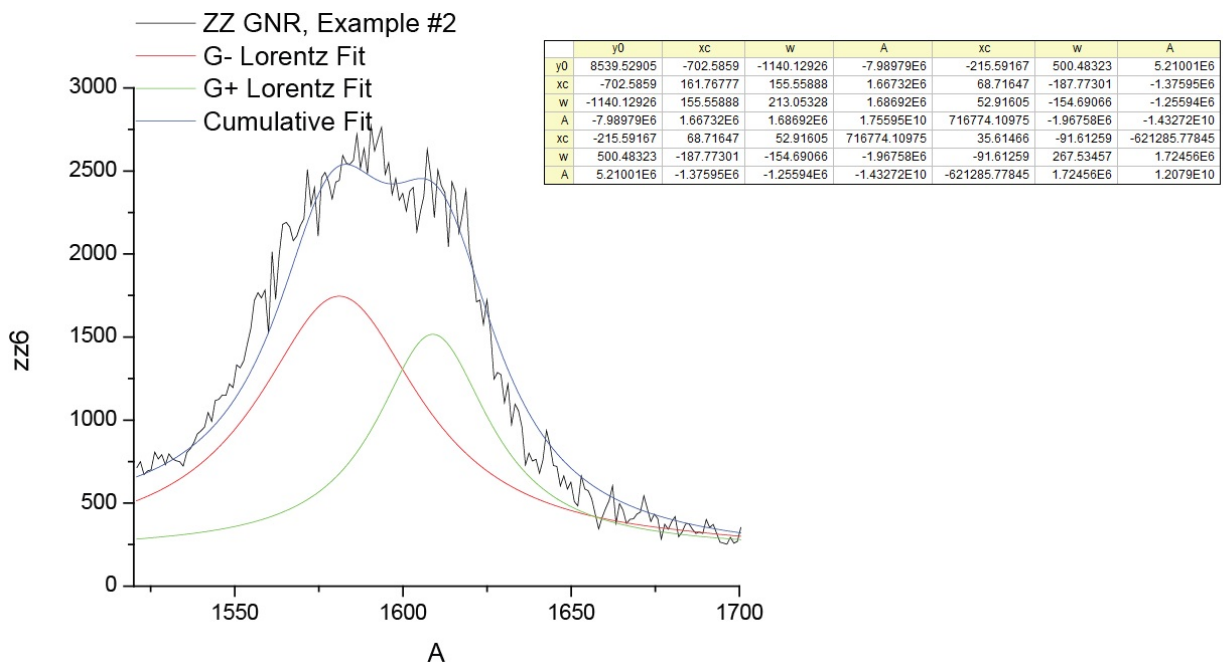

**Fig. 5-SM**. ZZ GNR example Raman spectrum (100 nm device) with two-peak Lorentzian fit. The covariance matrix may be seen at the top right corner of the graph.

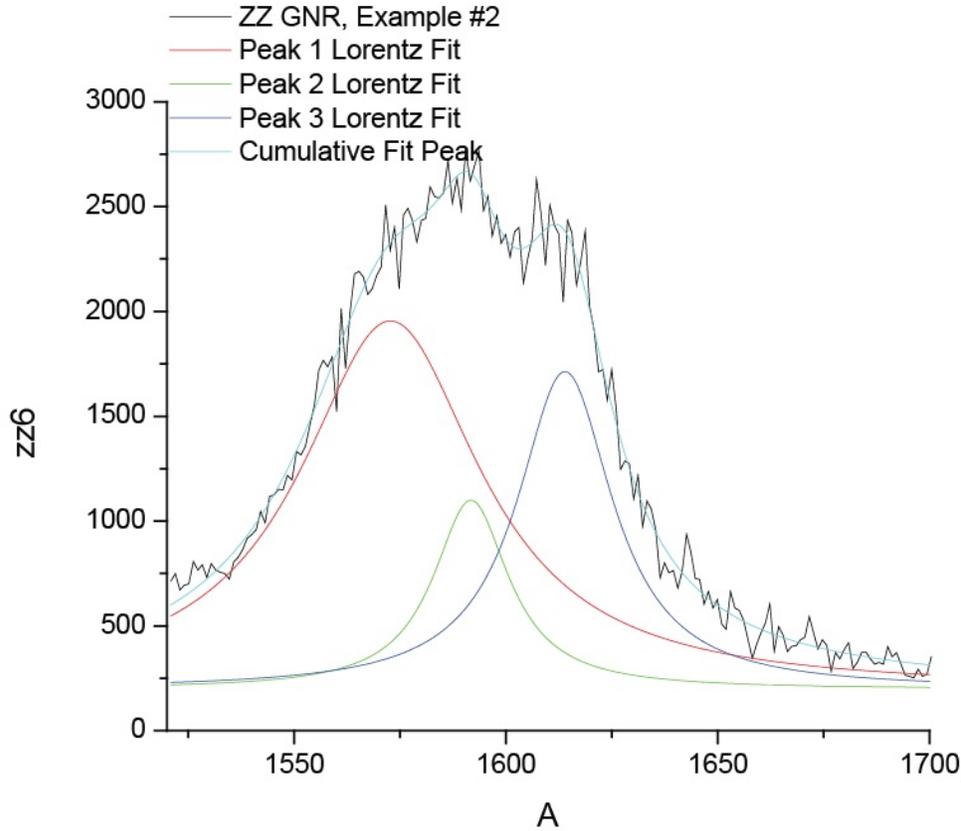

| | y0 | xc | w | A | xc | w | A | xc | w | A |
|---|---|---|---|---|---|---|---|---|---|---|
| y0 | 636.4852 | -33.79916 | -61.58332 | -382618.15644 | -4.92171 | 66.24134 | 218161.82353 | 3.68243 | -12.58134 | -24251.96986 |
| xc | -33.79916 | 4.62026 | 5.655 | 41952.50791 | 1.12865 | -8.59821 | -26920.29219 | -0.3401 | -0.24057 | -1833.62853 |
| w | -61.58332 | 5.655 | 10.29586 | 57509.99236 | 1.13892 | -8.32489 | -29759.09308 | -0.20987 | -0.48475 | -4665.4675 |
| A | -382618.15644 | 41952.50791 | 57509.99236 | 4.08236E8 | 9622.39661 | -83225.45498 | -2.56826E8 | -3586.58125 | 70.41077 | -9.27299E6 |
| xc | -4.92171 | 1.12865 | 1.13892 | 9622.39661 | 0.73762 | -1.37505 | -4829.90994 | 0.15131 | -0.59241 | -2531.94344 |
| w | 66.24134 | -8.59821 | -8.32489 | -83225.45498 | -1.37505 | 26.83897 | 69303.84222 | 1.90383 | -2.31031 | -10389.46289 |
| A | 218161.82353 | -26920.29219 | -29759.09308 | -2.56826E8 | -4829.90994 | 69303.84222 | 1.94584E8 | 4736.54643 | -5235.53074 | -1.8597E7 |
| xc | 3.68243 | -0.3401 | -0.20987 | -3586.58125 | 0.15131 | 1.90383 | 4736.54643 | 0.35069 | -0.59653 | -2350.98841 |
| w | -12.58134 | -0.24057 | -0.48475 | 70.41077 | -0.59241 | -2.31031 | -5235.53074 | -0.59653 | 2.60553 | 7944.10581 |
| A | -24251.96986 | -1833.62853 | -4665.4675 | -9.27299E6 | -2531.94344 | -10389.46289 | -1.8597E7 | -2350.98841 | 7944.10581 | 3.04792E7 |

**Fig. 6-SM**. ZZ GNR example Raman spectrum (100 nm device) with three-peak Lorentzian fit. The covariance matrix may be seen beneath the graph.

For the AC GNR example, several MLIs were calculated. For the one-peak model, $\sqrt{det\ \mathbf{Cov_p}}$ was calculated to be about $5 \times 10^5$, $n$ is 4, $L_{max}$ is about 1689, and $\prod_{i=1}^{n} \Delta p_i$ is on the order of $10^{11}$. In the two-peak model, $\sqrt{det\ \mathbf{Cov_p}}$ was calculated to be about $5 \times 10^{13}$, $n$ is 7, $L_{max}$ is about 1711, and $\prod_{i=1}^{n} \Delta p_i$ is on the order of $10^{20}$. In the three-peak model, $\sqrt{det\ \mathbf{Cov_p}}$ was calculated to be about $4 \times 10^{15}$, $n$ is 10, $L_{max}$ is about 1427, and $\prod_{i=1}^{n} \Delta p_i$ is on the order of $10^{29}$.

From these factors, the *logarithm* of the Bayes factor comparing the two-peak model to the one-peak model was greater than 5 (with the Bayes factor itself greater than 100), indicating that the two-peak model is a decisively stronger model to use. The *logarithm* of the Bayes factor comparing the three-peak model to the two-peak model was a positive number much smaller than 1 (and a logarithm on the order of –100), meaning that the three additional parameters introduced by the third peak were not appropriate from a statistical perspective.

The next analysis focuses on the 2D mode, which employs the Bayes factor calculation comparing a two-peak fit to a one-peak fit for AC and ZZ GNR cases in Fig. 7-SM and 8-SM, respectively. For the AC GNR one-peak model, $\sqrt{det\ \mathbf{Cov_p}}$ was calculated to be about $5 \times 10^4$, $n$ is 4, $L_{max}$ is about 2073, and $\prod_{i=1}^{n} \Delta p_i$ is on the order of $10^{11}$. In the AC GNR two-peak model, $\sqrt{det\ \mathbf{Cov_p}}$ was calculated to be about $2 \times 10^9$, $n$ is 7, $L_{max}$ is about 2065, and $\prod_{i=1}^{n} \Delta p_i$ is on the order of $10^{20}$. For the ZZ GNR one-peak model, $\sqrt{det\ \mathbf{Cov_p}}$ was calculated to be about $6 \times 10^4$, $n$ is 4, $L_{max}$ is about 2170, and $\prod_{i=1}^{n} \Delta p_i$ is on the order of $10^{11}$. In the ZZ GNR two-peak model, $\sqrt{det\ \mathbf{Cov_p}}$ was calculated to be about $10^8$, $n$ is 7, $L_{max}$ is about 1503, and $\prod_{i=1}^{n} \Delta p_i$ is on the order of $10^{20}$.

Calculating the MLIs and taking their ratio reveals decisively that the one-peak model is more appropriate than the two-peak model. Despite possible splitting reported in this mode, it is likely that the signal-to-noise ratio of the datasets is too low to objectively justify the use of a two-peak model.

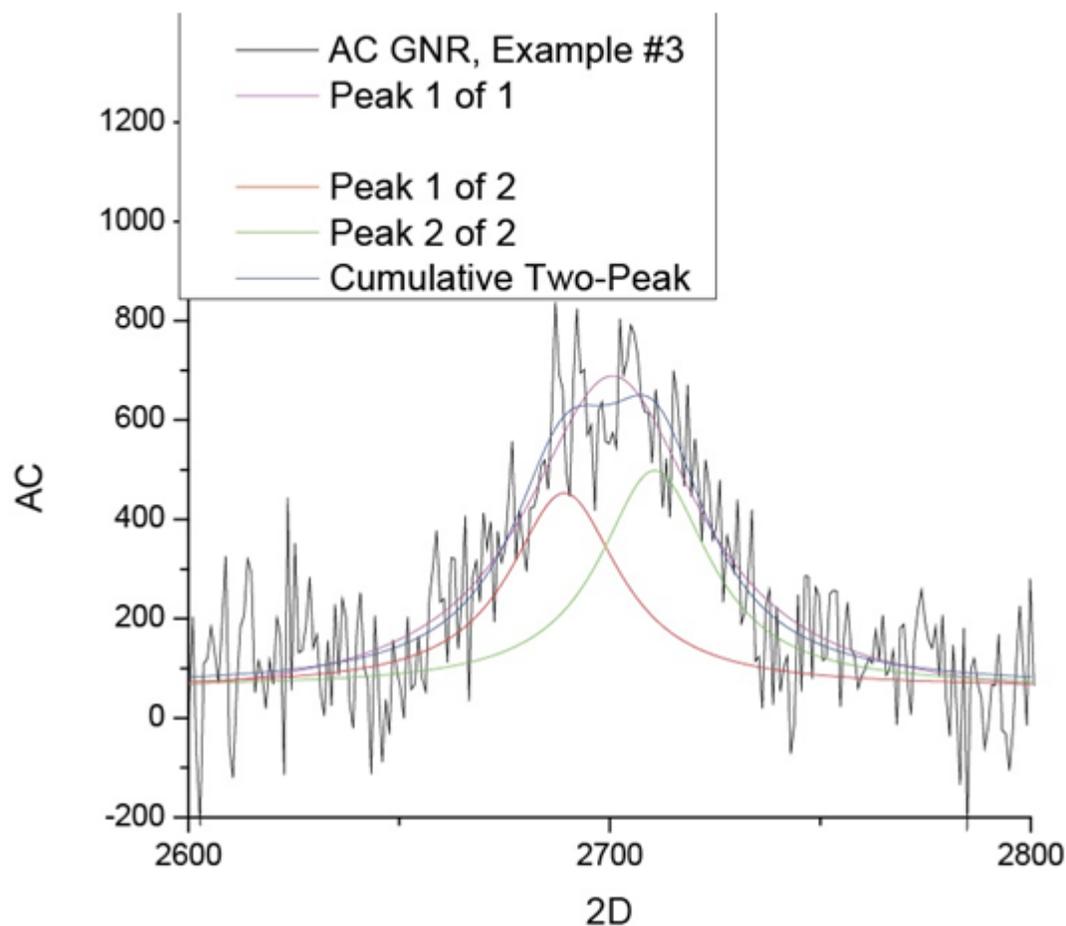

|   | y0 | xc | w | A |
|---|---|---|---|---|
| y0 | 298.71606 | 0.00596 | -53.53254 | -68106.80897 |
| xc | 0.00596 | 0.79454 | -0.01005 | -10.73875 |
| w | -53.53254 | -0.01005 | 16.27098 | 15726.48419 |
| A | -68106.80897 | -10.73875 | 15726.48419 | 1.91003E7 |

|   | y0 | xc | w | A | xc | w | A |
|---|---|---|---|---|---|---|---|
| y0 | 218.13902 | -0.76785 | -43.09553 | -20536.27423 | 1.24329 | -40.04067 | -22510.91382 |
| xc | -0.76785 | 11.85112 | 20.09999 | 27390.04286 | 8.97044 | -14.9674 | -26984.84494 |
| w | -43.09553 | 20.09999 | 61.03615 | 60613.28804 | 15.03732 | -22.18476 | -48467.45073 |
| A | -20536.27423 | 27390.04286 | 60613.28804 | 7.85012E7 | 24254.88658 | -43367.37403 | -7.30178E7 |
| xc | 1.24329 | 8.97044 | 15.03732 | 24254.88658 | 9.59856 | -16.24202 | -24724.71884 |
| w | -40.04067 | -14.9674 | -22.18476 | -43367.37403 | -16.24202 | 49.11968 | 54492.02021 |
| A | -22510.91382 | -26984.84494 | -48467.45073 | -7.30178E7 | -24724.71884 | 54492.02021 | 7.88384E7 |

**Fig. 7-SM**. AC GNR example Raman spectrum (100 nm device) with both the one-peak and two-peak Lorentzian models. Both covariance matrices may be seen beneath the graph.

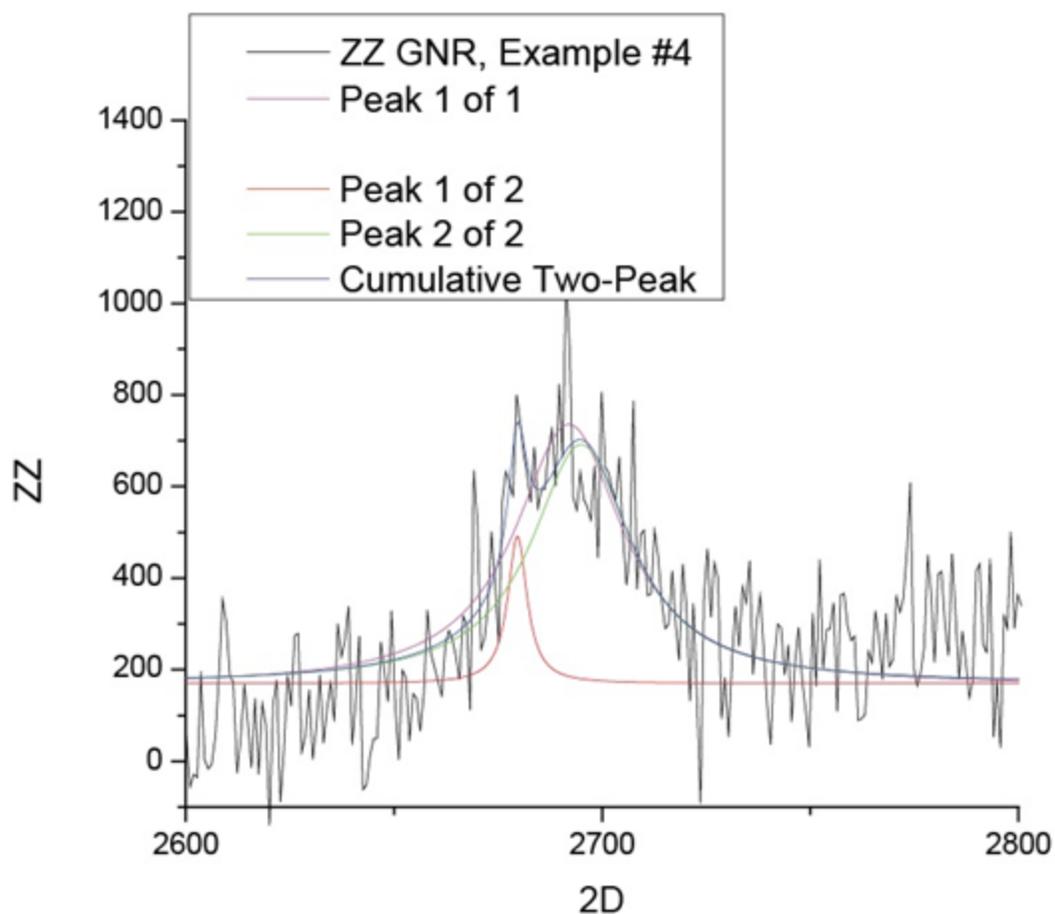

|    | y0         | xc       | w          | A           |
|----|------------|----------|------------|-------------|
| y0 | 237.05486  | -0.05698 | -37.78679  | -38383.9967 |
| xc | -0.05698   | 1.02733  | 1.88442E-5 | 1.54219     |
| w  | -37.78679  | 1.88442E-5 | 14.36043 | 9883.6685   |
| A  | -38383.9967 | 1.54219 | 9883.6685  | 9.57798E6   |

|    | y0          | xc        | w          | A         | xc        | w         | A           |
|----|-------------|-----------|------------|-----------|-----------|-----------|-------------|
| y0 | 246.66578   | 0.88311   | 14.03338   | 10034.71224 | 8.29883 | -54.64864 | -52426.00084 |
| xc | 0.88311     | 0.58271   | 0.34632    | 208.74473 | 0.24311   | -0.19122  | -435.04443  |
| w  | 14.03338    | 0.34632   | 9.87923    | 4540.85288 | 3.41652  | -7.64196  | -8547.82769 |
| A  | 10034.71224 | 208.74473 | 4540.85288 | 2.73162E6 | 2192.12789 | -5271.97406 | -5.56576E6 |
| xc | 8.29883     | 0.24311   | 3.41652    | 2192.12789 | 2.86905  | -4.31537  | -4538.66039 |
| w  | -54.64864   | -0.19122  | -7.64196   | -5271.97406 | -4.31537 | 25.11638 | 19798.2309  |
| A  | -52426.00084 | -435.04443 | -8547.82769 | -5.56576E6 | -4538.66039 | 19798.2309 | 1.91817E7 |

**Fig. 8-SM**. ZZ GNR example Raman spectrum (100 nm device) with both the one-peak and two-peak Lorentzian models. Both covariance matrices may be seen beneath the graph.

...

The final analysis focuses on the D mode, which again employs the Bayes factor calculation comparing a two-peak fit to a one-peak fit for AC and ZZ GNR cases in Fig. 9-SM and 10-SM, respectively. For the AC GNR one-peak model, $\sqrt{det\ \mathbf{Cov_p}}$ was calculated to be about $4 \times 10^4$, $n$ is 4, $L_{max}$ is about 1999, and $\prod_{i=1}^{n} \Delta p_i$ is on the order of $10^{11}$. In the AC GNR two-peak model, $\sqrt{det\ \mathbf{Cov_p}}$ was calculated to be about $10^{12}$, $n$ is 7, $L_{max}$ is about 1510, and $\prod_{i=1}^{n} \Delta p_i$ is on the order of $10^{20}$. For the ZZ GNR one-peak model, $\sqrt{det\ \mathbf{Cov_p}}$ was calculated to be about $2 \times 10^4$, $n$ is 4, $L_{max}$ is about 1986, and $\prod_{i=1}^{n} \Delta p_i$ is on the order of $10^{11}$. In the ZZ GNR two-peak model, $\sqrt{det\ \mathbf{Cov_p}}$ was calculated to be about $2 \times 10^{10}$, $n$ is 7, $L_{max}$ is about 1945, and $\prod_{i=1}^{n} \Delta p_i$ is on the order of $10^{20}$.

Calculating the MLIs and taking their ratio reveals that the one-peak model is, in most cases, decisively more appropriate than the two-peak model. It should be noted here that the D peak appears to have more ambiguity around its models. There are 2 out of 30 datasets that were calculated to have a logarithm of Bayes factor near unity (the two Bayes factors were about 0.3 and 9). The first consideration to make is that most datasets had Bayes logarithms of at least two digits, making their determinations unquestionable. The second consideration is that for the two "anomalous" values here, they fit the description of "substantial consideration" according to Dunstan et al. (see Ref. [56] of main text), but not "decisive". In other words, these two particular datasets are in a gray area of interpretation with no definitive answer on whether either model is clearly more appropriate. Given this information, it was scientifically conservative to continue adopting the one-peak model.

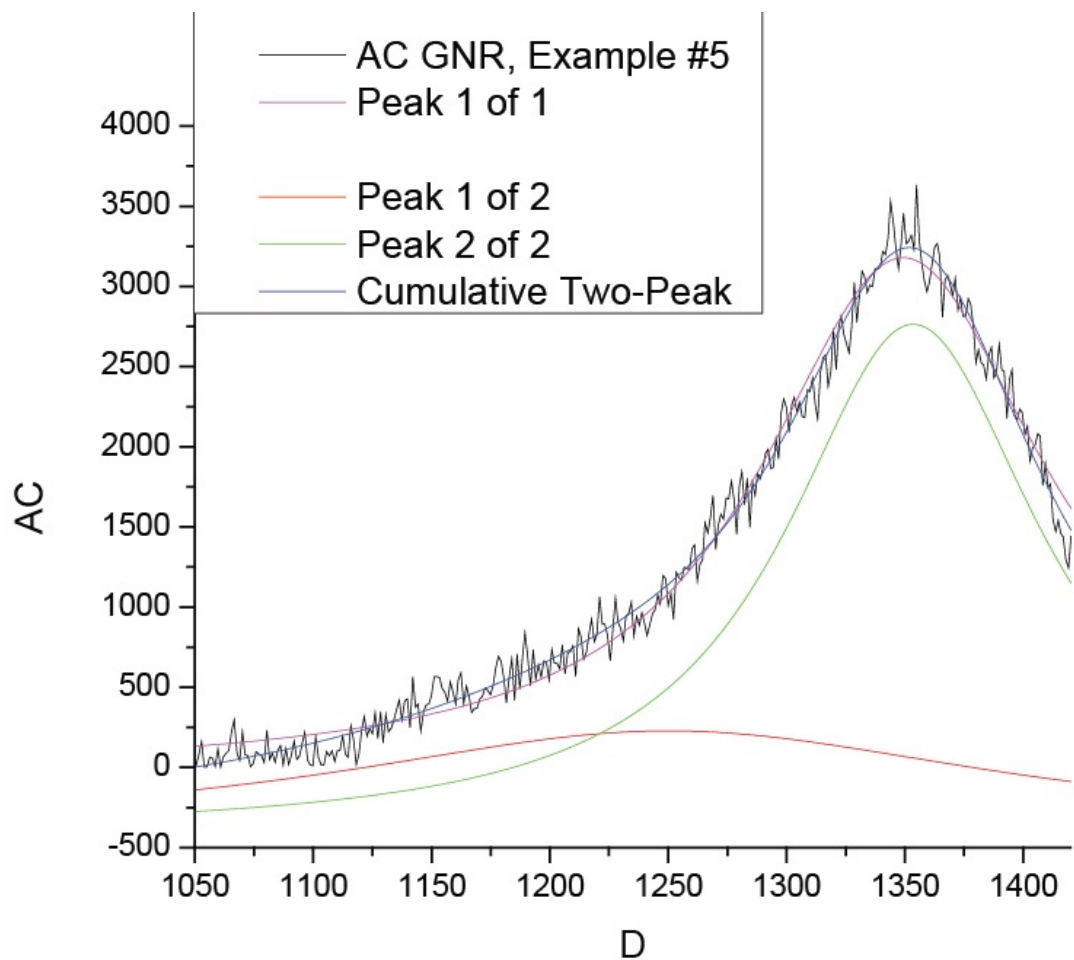

|    | y0          | xc       | w         | A            |
|----|-------------|----------|-----------|--------------|
| y0 | 319.54128   | 0.26455  | -29.63093 | -198078.61392|
| xc | 0.26455     | 0.22654  | 0.18508   | 669.3142     |
| w  | -29.63093   | 0.18508  | 5.14147   | 25801.23878  |
| A  | -198078.61392 | 669.3142 | 25801.23878 | 1.55909E8  |

|    | y0          | xc          | w           | A           | xc          | w          | A           |
|----|-------------|-------------|-------------|-------------|-------------|------------|-------------|
| y0 | 86161.27195 | -15594.80584 | -60622.41393 | -1.34845E8 | -115.16334  | 1982.1991  | 2.9083E7    |
| xc | -15594.80584 | 4042.72472 | 12258.09425 | 2.81168E7   | 31.20129    | -558.69118 | -7.68484E6  |
| w  | -60622.41393 | 12258.09425 | 44362.04655 | 9.89566E7  | 86.00932    | -1591.05316 | -2.29549E7 |
| A  | -1.34845E8  | 2.81168E7   | 9.89566E7   | 2.22534E11  | 210456.95432 | -3.72075E6 | -5.29575E10 |
| xc | -115.16334  | 31.20129    | 86.00932    | 210456.95432 | 0.50782    | -4.67327   | -61236.3632 |
| w  | 1982.1991   | -558.69118  | -1591.05316 | -3.72075E6  | -4.67327    | 82.81403   | 1.0872E6    |
| A  | 2.9083E7    | -7.68484E6  | -2.29549E7  | -5.29575E10 | -61236.3632 | 1.0872E6   | 1.47397E10  |

**Fig. 9-SM**. AC GNR example Raman spectrum (100 nm device) with both the one-peak and two-peak Lorentzian models. Both covariance matrices may be seen beneath the graph.

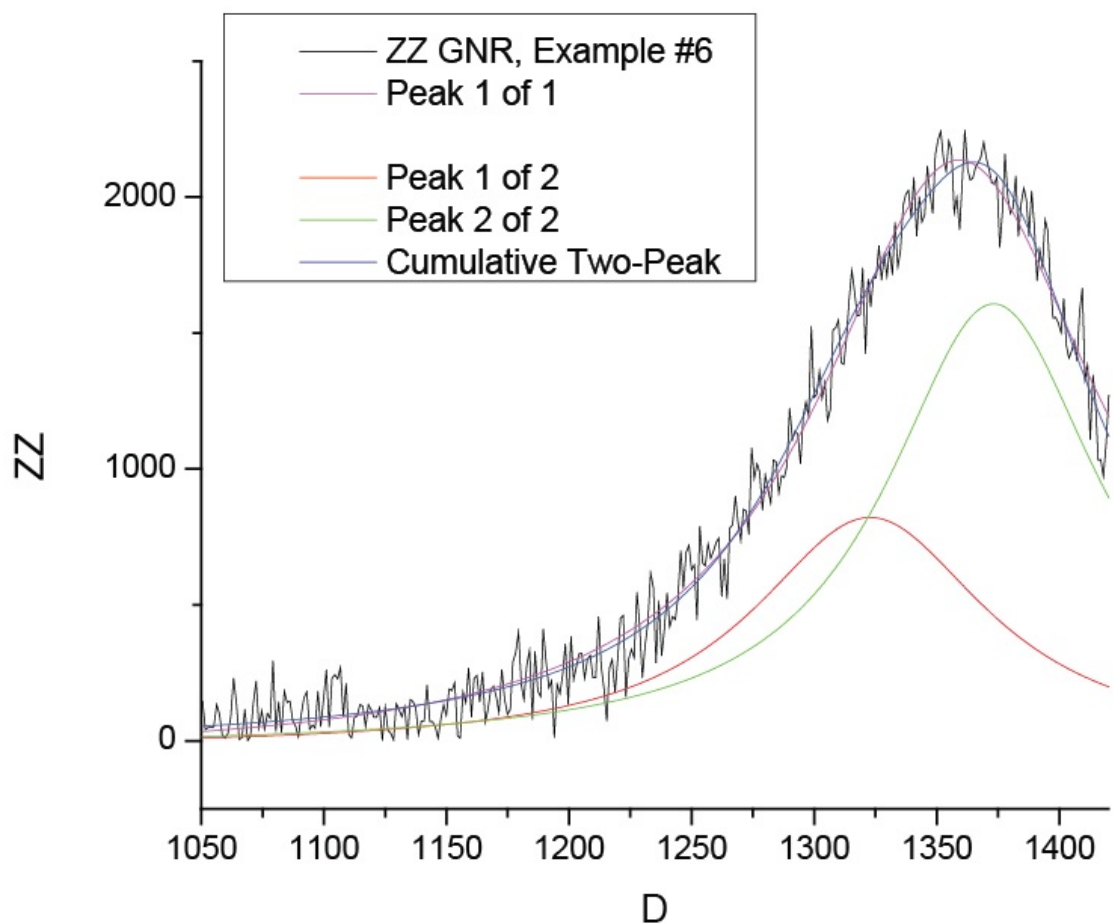

|    | y0           | xc         | w         | A             |
|----|--------------|------------|-----------|---------------|
| y0 | 163.7416     | -0.13318   | -22.97652 | -102572.44777 |
| xc | -0.13318     | 0.29724    | 0.34735   | 1008.16504    |
| w  | -22.97652    | 0.34735    | 6.26942   | 21082.34492   |
| A  | -102572.44777| 1008.16504 | 21082.34492 | 8.47255E7   |

|    | y0          | xc          | w           | A           | xc          | w          | A           |
|----|-------------|-------------|-------------|-------------|-------------|------------|-------------|
| y0 | 226.40496   | -55.91024   | -126.19382  | -426969.63004 | -16.74231 | 14.52414   | 294323.46585 |
| xc | -55.91024   | 90.82244    | 111.45579   | 664979.33908 | 39.72039  | -78.38834  | -655456.72096 |
| w  | -126.19382  | 111.45579   | 171.06763   | 837758.13634 | 45.01917  | -89.04156  | -782591.80224 |
| A  | -426969.63004 | 664979.33908 | 837758.13634 | 5.10584E9 | 307903.77849 | -645379.63989 | -5.08087E9 |
| xc | -16.74231   | 39.72039    | 45.01917    | 307903.77849 | 19.83754  | -41.25343  | -313853.09207 |
| w  | 14.52414    | -78.38834   | -89.04156   | -645379.63989 | -41.25343 | 101.21324  | 684573.72752 |
| A  | 294323.46585 | -655456.72096 | -782591.80224 | -5.08087E9 | -313853.09207 | 684573.72752 | 5.17321E9 |

**Fig. 10-SM**. ZZ GNR example Raman spectrum (100 nm device) with both the one-peak and two-peak Lorentzian models. Both covariance matrices may be seen beneath the graph.